\documentstyle[12pt]{article}
\newcommand{\ee}{$e^- e^+$}
\newcommand{\ep}{$e p$}
\newcommand{\ppb}{$p \bar{p}$}
\newcommand{\mH}{$M_{H}$}
\newcommand{\mW}{$M_{W}$}

\newcommand{\mt}{$M_{t}$}
\newcommand{\Z}{$Z^0$}
\newcommand{\pT}{$p_{T}$}
\newcommand{\Hbb}{$H  \to b \bar{b}$}
\newcommand{\Htata}{$H  \to \tau^+ \tau^-$}
\newcommand{\Hgaga}{$H  \to \gamma \gamma$}

\newcommand{\HZZs}{$H\to Z^0 Z^*$}

\newcommand{\ggH}{$g g \to  H $}
\newcommand{\ppWH}{$p p \to W H X $}
\newcommand{\ppZH}{$p p \to Z^0 H X $}
\newcommand{\ppttH}{$p p \to t \bar{t} H X $}
\newcommand{\ppttZ}{$p p \to t \bar{t} Z^0 X$}
\newcommand{\ppttW}{$p p \to t \bar{t} W X $}
\newcommand{\ppWWW}{$p p \to W W W X $}
\newcommand{\ppWWZ}{$p p \to W W Z^0 X $}

\newcommand{\ppWWj}{$p p \to W W j  X $}

\newcommand{\ppttj}{$p p \to t \bar{t} j X$}

\newcommand{\ppWWjj}{$p p \to W W j j X $}

\newcommand{\ppttjj}{$p p \to t \bar{t} j j X$}

\newcommand{\ppWZ}{$p p \to W Z^0 X $}
\newcommand{\ppZZ}{$p p \to Z^0 Z^0 X $}

\newcommand{\ppWjjj}{$p p \to W j j j X$}

\newcommand{\ppZj}{$p p \to Z^0 j X$}

\newcommand{\pptt}{$p p \to t \bar{t} X $}

\newcommand{\Wlepton}{$ W \to \ell \nu_\ell $}
\newcommand{\Zlepton}{$ Z \to \ell^+ \ell^- $ }

\newcommand{\telmu}{$ t \to b  e(\mu)  \nu $}
\newcommand{\tauhad}{$ \tau \to hadrons + \nu_\tau$}

\newcommand{\sigd}{ ``isolated $e \mu$ and a tau-jet + missing $p_T$''}
\newcommand{\siga}{ ``isolated $e \mu$ and two tau-jets + missing $p_T$''}
\newcommand{\sige}{ ``isolated $e (\mu)$ pair and a  tau-jet + missing $p_T$''}
\newcommand{\sigc}{ ``isolated $e (\mu)$ and three  tau-jet + missing $p_T$''}
\newcommand{\sigb}{ ``isolated $e (\mu)$ pair and two tau-jets + missing $p_T$''}

\newcommand{\sigf}{ ``isolated $e (\mu)$ and two  tau-jet + $\geq 3$ jets+ missing $p_T$''}
\newcommand{\sigg}{ ``isolated three tau-jet + $\geq 3$ jets+ missing $p_T$''}
\newcommand{\sigh}{ ``isolated $e (\mu)$ and a  tau-jet + $\geq 3$ jets+ missing $p_T$''}
\newcommand{\sigi}{ ``isolated two tau-jet + $\geq 5$ jets+ missing $p_T$''}
\newcommand{\sigj}{ ``isolated $e (\mu)$ and a  tau-jet + a bottom jet + missing $p_T$''}
\newcommand{\sigk}{ ``isolated tau-jet + 6 jets + missing $p_T$''}

\newcommand{\Mtata}{$M(\tau \tau)$}

\newcommand{\beq}{\begin{equation}}
\newcommand{\eeq}{\end{equation}}
\newcommand{\bea}{\begin{eqnarray}}
\newcommand{\eea}{\end{eqnarray}}

\def\ie{\hbox{\it i.e.}{}}    
    
\def\viz{\hbox{\it viz.}{}}
\def\etal{\hbox{\it et al.}{}}
\relax

\def\figcap{\section*{Figure Captions\markboth
     {FIGURECAPTIONS}{FIGURECAPTIONS}}\list
     {Fig. \arabic{enumi}:\hfill}{\settowidth\labelwidth{Fig. 999:}
     \leftmargin\labelwidth
     \advance\leftmargin\labelsep\usecounter{enumi}}}
 \relax
\def\reflist{\section*{REFERENCES\markboth
     {REFLIST}{REFLIST}}\list
     {[\arabic{enumi}]\hfill}{\settowidth\labelwidth{[999]}
     \leftmargin\labelwidth
     \advance\leftmargin\labelsep\usecounter{enumi}}}
 \relax
\def\tabcap{\section*{Tables\markboth
     {TABLES}{TABLES}}\list
     {Table \arabic{enumi}:\hfill}{\settowidth\labelwidth{Table 999:}
     \leftmargin\labelwidth
     \advance\leftmargin\labelsep\usecounter{enumi}}}
 \relax

\begin{document}
\begin{titlepage}
 \null
 \vskip 0.5in
\begin{center}
\makebox[\textwidth][r]{IP/BBSR/98-6}

 \vspace{.15in}
  {\Large
     MORE SIGNATURES OF THE INTERMEDIATE MASS HIGGS BOSON
    }
  \par
 \vskip 1.5em
 {\large
  \begin{tabular}[t]{c}
    Pankaj Agrawal  \\
\em Institute of Physics \\
\em Sachivalaya Marg \\
 \em Bhubaneswar, Orissa 751005 India\\
  \end{tabular}}
 \par \vskip 5.0em
 {\large\bf Abstract}
\end{center}
\quotation

  We examine the potential signatures of the Higgs boson when
  it decays into a tau-lepton pair.  We show that with the proper
 identification of the tau-jet, this decay mode 
 can lead to the identification of the Higgs
 boson over most of the intermediate mass range (\mW $<$ \mH $<$ 2 \mW).

\endquotation
\baselineskip 10truept plus 0.2truept minus 0.2truept 
\vfill 
\mbox{Jan 1998} 
\end{titlepage} 
\baselineskip=21truept plus 0.2truept minus 0.2truept 
\pagestyle{plain}
\pagenumbering{arabic}


   Recent experiments at the LEP (a \ee~collider), the Tevatron (a \ppb~collider), 
   the HERA (a \ep~collider) machines, and a host of other
   experiments have tested the standard model in quite diverse situations.
   Still there is no established disagreement with the model. However one
     important part of the standard model, \viz, the Higgs mechanism, has 
     not been yet validated. One of the important consequences  of this mechanism
     is the existence of a neutral scalar boson, the Higgs boson.
     The existence of the Higgs boson was predicted more than 30 years ago
     and there have been active searches for this particle. More recently
     experiments at the LEP machine have set a lower bound on the mass of the Higgs 
     boson, \mH $> O(80)$ GeV [\ref{lep1}]. This lower bound can go as high as $90-100$
     GeV in coming years. Beyond this mass range, the Higgs boson can be
     searched for at the LHC machine. (If high luminosity option at the Tevatron,
     TeV33, is realized, then the Higgs boson up to a mass of $130-140$ GeV
     can also be searched at this machine.) The CMS and ATLAS 
     collaborations are optimizing their detectors
     at the LHC machine so as to be able to identify specific
     Higgs boson signatures. In this letter, we focus on the 
     intermediate mass range, \mW $<$ \mH $< 2$ \mW, of the Higgs boson
     and identify some more signatures of this particle.

     The popular signature in the lower intermediate mass range 
     (\mW $<$ \mH $< 130-140$ GeV) has been
     ``isolated two photons'' [\ref{lhc1}]. This signal arises when the Higgs boson
     decays into two photons (\Hgaga), after being produced through
     gluon annihilation, \ggH. (This process is at the lowest order;
     higher order processes are also important.) This decay mode of
     the Higgs boson is a rare decay channel with a branching ratio
     of the order of $10^{-3}$ in the relevant mass range. The CMS collaboration
     detector at the LHC machine is specifically designed so that it could
     identify a photon with sufficient precision and overcome the
     background of the direct two photon production. If design precisions are 
     achieved, this signature will have a good significance (Here significance 
     is defined as $S\over\sqrt{B}$, where S is the number of signal events and 
     B is the number of background events.), 
     but poor signal-to-background ratio. There has also been at least
     one other signature, ``exclusive isolated $e/\mu + 2$ bottom jets'',
     that have been advocated in the past. Leading order parton level
     calculations [\ref{ae},\ref{smw},\ref{abc}] have shown that this signature can 
     be useful. But,
     a study [\ref{fr}] based on event generators has cast doubt on the utility
     of this signature. However, as has been pointed out [\ref{abc},\ref{ak}], 
     an appropriate
     modification of this signature by including an extra `soft' jet,
     \ie, ``exclusive isolated $e/\mu + 2$ bottom jets + a `soft' jet'',
     can still be useful. This signature is based on the dominant
     decay mode \Hbb~in the relevant mass range. In the upper
     intermediate mass region ($130-140$ GeV $<$ \mH~$<$ 2 \mW), the
     decay mode \HZZs~can provide useful signatures [\ref{hunter}]. In this letter, we
     suggest some other possible useful signatures in the 
     intermediate mass range which are based on the decay mode \Htata,
     after the Higgs boson has been produced through the process \ppttH.

       The decay mode \Htata~is sub-dominant in the intermediate
       mass range. But there are at least two features
       of this decay mode that are in its favor: 1) The decay
       products are leptons, so there will be less of a strong
       interaction background; 2) the branching ratio is not
       very small, about $5\%$ in the lower intermediate mass
       region (decreasing sharply only after about \mH $> 150$ GeV,
       as \mH~approaches the $WW$ threshold).
        At the same time there is at least one feature
       of this decay mode that is not in its favor, \viz,
       a tau-lepton always has a neutrino in its decay products.
       Therefore one cannot fully reconstruct the four-momentum
       of the tau-lepton; which in turn means there is no
       resonance in the \Mtata~distribution. (One can attempt to 
       reconstruct the four-momentum of the tau-lepton
       in this decay mode by using the fact that tau-lepton
       has a small mass and will have large momentum.)
       However, as we shall see, one can use this decay
       mode without reconstructing \Mtata. The largest
       event rate for this channel occurs when the Higgs
       boson is produced through gluon annihilation, \ggH.
       But due to large backgrounds, this is not a useful
       mechanism. The usefulness of other production 
       mechanism, \ppZH~and \ppWH~for this decay mode are
       not yet fully established. Our focus here will
       be on the production mechanism \ppttH.

	 A tau-lepton can decay purely leptonically, or
	 semi-leptonically, \ie , into hadrons and a neutrino. When a 
	 tau-lepton decays semileptonically, at large \pT, it 
	 manifests itself as a jet, a tau-jet. This tau-jet differs in 
	 important ways from a parton-initiated jet. The tau-lepton
	 has a small mass and is color singlet. Because of this
	 a tau-jet is a narrow stream of at most a few pions. Low
	 multiplicity and the narrowness sets a tau-jet apart from a 
	 parton-initiated jet, specially at the LHC where a parton-initiated
	 jet will have a very high multiplicity of hadrons in it.
	 Tau-jets have been identified at both \ee~and hadron 
	 colliders [\ref{cdf1}]. It can be hoped that at the LHC, a tau-jet
	 can be identified with a very good precision.
	 Still, there will be some probability for a few hadrons
	 to stray away from a parton-initiated jet and mimic a
	 tau-jet. However probability of such faking will certainly
	 be lower than that of a jet faking a bottom jet. This
	 later probability is about a percent. Therefore in this
	 study, we are assuming that a tau-jet can be identified
	 at the LHC with a contamination level of less than a
	 percent.

	 The production mechanism, \ppttH,  with the subsequent decay
	 \Htata~can lead to a number of useful signatures.
	 These different signatures come about depending on:
	 a) what decay modes of the top quark we look at for the
	 top quark and the anti-top quark; b) what particles we identify 
	 in the final state; c) whether we tag a bottom jet or
	 not. More isolated leptons in the signature would mean
	 less of a background. An isolated lepton in the signature
	 means a $W/Z$ boson in the background (directly produced
	 or through the top quark decay for the $W$ boson), which
	 tends to reduce the background. Also minimizing the 
	 number of jets in the signature leads to a smaller
	 background, because jets can originate through strong
	 interaction processes. In brief, the guiding philosophy
	 to choose a good signature is: more of leptons and less
	 of jets. We hasten to add that this philosophy can
	 lead to smaller number of events of the signal. So
	 sometimes one has to strike a balance. Below we present
	 a few signatures of the Higgs boson in the 
	 intermediate mass range. Some of these signatures are
	 analyzed in more detail than the other.

   1) {\bf Four-lepton signatures}: 
   
      These signatures have electron, muon
      and tau-leptons in it. As discussed above, the tau-lepton
      will manifest itself as a tau-jet. We will enumerate these
      signatures below. These signatures occur when the Higgs
   boson is produced with a pair of top quarks (\ppttH) and subsequently
   both top quarks decay semi-leptonically and the Higgs boson
   decays into a tau-lepton pair. Afterwards tau-leptons decay
   into hadrons and neutrino, thus giving rise to tau-jets.
   All of these signatures will have \ppttZ~as the major background.
   (The process $pp \to t \bar{t} \gamma^{*} X (\gamma^{*} \to \tau \tau)$
    and its interference with \ppttZ~(with the \Z~decay) will also
    be a background. However a \pT~cut of order $15$ GeV on a tau-jet,
    which we shall be making, will suppress it relative to the
    \ppttZ~process.)
   We shall call such backgrounds as direct-backgrounds. As we
   discussed earlier, there is also a probability of a
   parton-initiated jet to fake a tau-jet, therefore we shall
   have to consider some additional backgrounds. We shall call
   such backgrounds mimic-backgrounds. Since
   four-lepton signatures have a pair of tau-jets, the 
   major mimic-backgrounds will be \ppttjj~and \ppWWjj. However,
   we shall assume that this mimic-probability is less than a
   percent, and thus ignore these mimic-backgrounds. We would
   note that even if mimic-probability is a few percent, these
   backgrounds would not be significantly larger than the signal.
   (The processes \pptt~and \ppttj~can also be backgrounds when 
   both the top quarks decay semi-leptonically and the resulting 
   bottom jets mimic tau jets. We are ignoring these backgrounds 
   assuming such mimic probability to be quite small. In any case, 
   the signal will have accompanying jets. Observation
   of even one of them will reduce the backgrounds further.)
   Let us now discuss various four-lepton signatures.

   1a) {\em \siga}: This signature occurs when one of the top quarks
   decays into an electron (and other particles), while the other
   top quark decays into a muon (and other particles). (For the
   purpose of our discussion, the distinction between the
   top quark and anti-top quark is immaterial; therefore to
   avoid oblique language, we shall refer to top quark. It will
   be understood that this top quark could be an anti-top
   quark.) After \ppttZ, the next significant direct background 
   will be \ppWWZ; this will be a background when the two $W$-bosons
   decay into $e(\mu)$ and neutrino, while the \Z-boson decays into a 
   tau-lepton pair.  However, contribution of this process
   to the signature will be about an order of magnitude smaller
   than the signal contribution. This background is smaller,
   because this is a electroweak process.
   In Table 1, we present contribution of the signal and
   the major background \ppttZ~to this signature..
    We have used PYTHIA 5.7 [\ref{pythia}] to compute with basic
   cuts on the observables. Without cuts we have compared the
   PYTHIA results with a parton level calculation of \ppttH
   and \ppttZ. We find reasonable agreement. We have applied
   following basic cuts:

   \beq
   p_{T}^{e, \mu, \tau} > 15 \;\; {\rm GeV}; \;\;\;\;\;\;   |\eta|^{e, \mu, \tau} < 2.5; 
   \;\;\;\;\;\;  \Delta R(\ell, \ell^\prime) > 0.6.
   \eeq

   Here the index $\tau$ stands for tau-jet not the tau-lepton. $\eta$ is
   pseudo-rapidity and 
   $\Delta R(\ell,\ell^\prime)=\sqrt{(\Delta \eta)^2 + (\Delta \phi)^2 }$.

   In running PYTHIA, we have used its proper handling of the tau-lepton
   decay. We have not used its full functionality. This is because we
   are looking at inclusive leptonic signatures. Initial and final state
   radiations as well as hadronization should not affect our results in
   any significant way. 

     At the LHC machine ($\sqrt{s} = 14$ TeV), in the mass range \mH~= $80-140$ GeV
     and \mt~$= 175$ GeV, 
     without any cuts or branching ratios, 
     the \ppttH~cross-section varies between $2.1-0.41$ pb. The similar cross-section
     for the \ppttZ~is about $0.45$ pb. The \Htata~branching ratio varies from 5.6 to 3.1\%.
     The numbers in the table are for integrated luminosity of $10^5$ pb$^{-1}$. We see
     that as we go from the Higgs boson mass of 80 to 140 GeV, effect of
     the cuts decreases. This is simply because, as \mH~increases, the \pT~of
     the tau-lepton gets harder; so does the \pT~of the tau-jets. Because of this
     harder \pT~distribution, the reduction in the event rate gets smaller.
     We also note that the cross-section for the \ppttZ~process is not large
     compared to the \ppttH~process (in the intermediate mass region of the
     Higgs boson). This happens because the $t\bar{t}H$ coupling is proportional
     to $M_{t}\over M_{W}$ (\mt~= 175 GeV and \mW~= 80 GeV) and the mass of the Higgs
     boson is of same order as the $Z$-boson.

   1b) {\em \sigb}: This signature comes to be when both top quarks of
   the signal decay into electrons or muons. In this case,
   compared to the signature 1a), there is one additional 
   direct-background: \ppZZ. Without any suppression cuts, this
   will be the largest direct-background. However, the mass-distribution
   of the electron/muon pair from the \Z~decay will be sharply peaked 
   at the mass of the $Z$-boson; while this distribution for
   the signal will be quite diffused because the two electrons/muons
   will come from the decay of different top quarks. Therefore a
   cut in this distribution, can eliminate this background. We
   are then left with the same backgrounds as in the case of the
   signature 1a). In the Table 1, we also give results for this
   signature, with the same cuts, as the signature 1a). This is
   basically just a multiplication by a factor of $2$.
   We see that naive significance varies about from $25$ to $4$. We have 
   not listed them in the table, because these numbers only indicate the
   usefulness of the signatures. Actual values may be lower if
   the mimic-probability is not less than one percent. Even with few
   percent mimic-probability, we can have reasonably good significance
   for the lower intermediate mass region. We also notice that
   the accumulated integrated luminosity of the order of $3-5 \times
   10^{4}$ may even be good enough for the detection.

   1c) {\em \sigc}: This signature occurs when, one of the top quark
      of the signal decays into a tau-lepton (and other particles),
      while the other top quark decays into an electron/muon
      (and other particles).
      The number of signal events in this background will be 
      smaller by at-least a factor one-third (compared to 
      that in signatures 1a) and 1b)). This is because of
      the branching ratio for \tauhad~is about $64\%$. In addition
      there will be a new mimic-background: \ppWjjj. Both of
      these potential problems occur because now we have a 
      third tau-jet in the signature. The usefulness of
      this signature will depend on how well one can discriminate
      a tau-jet from a parton-initiated jet.

   2) {\bf Three-lepton Signatures}:

       These signatures will have one tau-jet and the other two
       leptons can be either both electrons/muons or one can
       be electron and the other muon. (The three-lepton signature
       with more than one tau-lepton will suffer from the large 
       mimic-background of \ppZj.) In the case of this signature we
       observe only one tau-lepton, while the second tau-lepton
       from the Higgs boson decay is left unobserved. The usefulness
       of three-lepton signatures lies in having larger signal
       events, as compared to the four-lepton signatures. This happens
       because we do not loose in observing second tau-jet in terms
       of \tauhad~branching ratio and do not have to apply cuts 
       (so have more phase space available). The major direct-backgrounds
       for this signature are: \ppttZ~and \ppttW. The contribution
       of the \ppttZ~will increase, because the top quarks
       and Z-boson can now decay in many more ways to three-leptons.
       However, it will only be about a factor of two larger than
       the signal. The contribution of the \ppttW~will be of same
       level as the \ppttZ~background. This is because the cross-section
       of \ppttW~at the LHC machine is about a factor of three smaller
       than the \ppttZ~process [\ref{bsp}], but the branching ratio \Zlepton
       is about a factor of three smaller than \Wlepton. (The
       cross-section of \ppttW~is smaller than \ppttZ, because the latter
       process can occur through gluon annihilation, but the former
       cannot.) One additional background, \ppWWW, is again quite small,
       because of being an electroweak process. Despite all this,
       the direct-backgrounds will be manageable, because we have
       more signal events and we have to take square root of the
       background to compute the naive significance. The major
       threat to this signature could be the mimic-backgrounds:
       \ppttj~and \ppWWj. Therefore, unless the mimic-probability
       of a parton-initiated jet mimicking a tau-jet can be
       sufficiently small, of the order of $10^{-4}$ or smaller,
       there could be problems with this signature. If the
       mimic-probability is sufficiently small, then the potentially
       useful signatures can be: \sigd~and \sige. The \sige~would 
       appear to have a large background from \ppWZ~process;
       however, as in the case of signatures 1b), a cut on the
       di-lepton mass, for the electrons/muons from the Z-boson
       decay, can eliminate this background.

    3) {\bf Three-lepton and multijet}:

         We can increase the number of signal events by observing
	 that class of events where one of the top quarks in
	 the process \ppttH~decays hadronically, \ie, into three jets.
         While the branching ratio for \telmu~is $2\over9$, for 
	 the decay $t \to 3$ jets it is $6\over9$. So in principle,
	 we could have a gain of a factor of 3 in the number of signal
	 events over that the signature 2 and even more over the four-lepton
	 signatures. In reality, because of identifying experimental
	 cuts, the gain in the signal events will be somewhat smaller.
	 The background events will also be correspondingly higher,
	 because now we can have strong interaction production of
	 jets also. Below we discuss some of the signatures of this class.

      3a) {\em \sigf}:  The largest direct background will be from
	  $pp \to W Z^0 X + \geq 3$ jets. As in case of the top-quark
	  searches, these backgrounds would be manageable.
	  Tagging one or more bottom jets will help further.
	  However the mimic-backgrounds can be significant.
	  These will be due to: \pptt, \ppttj, $pp \to W + \geq 5$ jets.
	  Low mimic-probability will be important for the
	  suppression of these backgrounds.

      3b) {\em \sigg}: Here backgrounds will be similar to that of
	  signature 3a). However, as discussed above,
	  we shall have smaller signal events, because of one
	  extra tau-jet identification.

    4) {\bf Two-lepton and multijet}: 
    
	As discussed in the case
	of signature 2, we can enhance the signal by identifying
	smaller number of leptons. The signal will have same origin as
	that in the signature 3, except we detect only one of the 
	tau-lepton that the Higgs boson decays into. 
	Some of the possible signatures in this category are:
	\sigh,\sigi,\sigj. However all of them are likely to suffer from
	the large $pp \to t \bar{t} + \geq 0$ jet background, which
	is one to two orders of magnitude higher than the signal.
	Because of very similar topology of the signal and the background,
	it may also be difficult to device cuts to suppress the
	background.

    5) {\bf One-lepton and multijet}:

         When both top quarks of the process \ppttH decays hadronically,
	 and we observe only one of the tau-lepton into which the
	 Higgs boson decays, we would get this signature. There will
	 be largest signal events in this signature (of all the signatures
	 that we have discussed). This signature, \sigk, will have direct-background
	 from $pp \to t \bar{t} + 2$ jets and $pp \to W + 6$ jets and 
	 the mimic-background from $pp \to 7$ jets. It may be difficult
	 to separate the signal from the background.

     In conclusion, we find that, at the LHC machine,  one should be able to observe the
     Higgs boson, if it exists and is in the intermediate mass range, through
     `four-lepton' and `three-lepton + multijet' signatures. 
     In particular, signatures \siga~and \sigb~appear to be quite useful. With
     enough accumulated luminosity, at the LHC machine, one should be able to search
     for the Higgs boson in the mass region of $80-150$ GeV. In some classes
     of the supersymmetric models, a scalar boson of properties similar
     to the standard model exists. Same techniques can be used to identify this
     scalar particle also.

\vskip .2in

This work was completed at the 5th Workshop on High Energy Physics 
Phenomenology (WHEPP-5) which was held in IUCAA, Pune, India. I thank
the organisors of the workshop.

\vskip .5in

\relax
\def\pl#1#2#3{
     {\it Phys.~Lett.~}{\bf B#1} (19#3) #2}
\def\zp#1#2#3{
     {\it Zeit.~Phys.~}{\bf C#1} (19#3) #2}
\def\prl#1#2#3{
     {\it Phys.~Rev.~Lett.~}{\bf #1} (19#3) #2}
\def\rmp#1#2#3{
     {\it Rev.~Mod.~Phys.~}{\bf #1} (19#3) #2}
\def\prep#1#2#3{
     {\it Phys.~Rep.~}{\bf #1} (19#3) #2}
\def\pr#1#2#3{
     {\it Phys.~Rev.~ }{\bf D#1} (19#3) #2}
\def\np#1#2#3{
     {\it Nucl.~Phys.~}{\bf B#1} (19#3) #2}
\def\ib#1#2#3{
     {\it ibid.~}{\bf #1} (19#3) #2}
\def\nat#1#2#3{
     {\it Nature (London) }{\bf #1} (19#3) #2}
\def\ap#1#2#3{
     {\it Ann.~Phys.~(NY) }{\bf #1} (19#3) #2}
\def\sj#1#2#3{
     {\it Sov.~J.~Nucl.~Phys.~}{\bf #1} (19#3) #2}
\def\ar#1#2#3{
     {\it Ann.~Rev.~Nucl.~Part.~Sci.~}{\bf #1} (19#3) #2}
\def\ijmp#1#2#3{
     {\it Int.~J.~Mod.~Phys.~}{\bf #1} (19#3) #2}
\def\cpc#1#2#3{
     {\it Computer Physics Commun. }{\bf #1} (19#3) #2}
\begin{reflist}

\item \label{lep1} S. Banerjee, Talk presented at WHEPP5, Pune, India (Janaury 1998).

\item \label{lhc1} ATLAS Letter of Intent, CERN/LHCC  92-4 (October 1992); CMS Letter of Intent, CERN/LHCC 92-3 (October 1992).

\item \label{ae} P. Agrawal and S. Ellis, \pl{229}{145}{89}.

\item \label{smw} A. Stange, W. Marciano, and S. Willenbrock,
\pr{49}{1354 }{94}; \pr{50}{4491}{94}.

\item \label{abc} P. Agrawal, D. Bowser-Chao and  K. Cheung, \pr{51}{6114}{95}.

\item \label{fr} D. Froidevaux and E. Richter-Was, preprint CERN-TH 7459/94.

\item \label{ak} P. Agrawal and M. Kar, IOP preprint IP/BBSR/97-31.

\item \label{hunter} J.F.~Gunion \etal, The Higgs Hunter's Guide,
Addison-Wesley  (1990).

\item \label{cdf1} The CDF Collaboration, F. Abe \etal, \prl{79}{3585}{97}.

\item \label{pythia}  H.-U. Bengtsson and T. Sjostrand, 
	 {\it Computer Physics Commun.} 46 (1987) 43.

\item \label{bsp} V. Barger, A. Stange and R. J. N. Phillips, \pr{45}{1484}{92}.

\end{reflist}

\newpage


\vskip .1in
\hskip -.5in

\begin{tabular}{||c|r|r|r|r|c|c||} \hline
 \mH  & \multicolumn{2}{c|}{Signal 1a)}  & \multicolumn{2}{c|}{Background}  & Signal 1a) + 1b)   &  Total  \\ \cline{2-5}
 (GeV) & no cuts &  with cuts & no cuts & with cuts & with cuts  & Background \\ \hline
 80 & 119  &   41 &  15 & 5  & 82 & 10  \\
 100 &  67  &   27 & 15 &  5 &  54  & 10  \\
 120 &  36  &   16 & 15 & 5 & 32 & 10 \\
 140 &  13  &    7 & 15 &  5 & 14 & 10 \\ \hline
\end{tabular}
\vspace{.2in}

{\small
Table 1.  The signal and the background rates at the LHC machine energy,
$\sqrt{s} = 14$ TeV, with the integrated luminosity of $10^{5}$ pb$^{-1}$
without any cuts and the cuts specified in the text. These for the
signatures 1a), \siga, and 1b), \sigb.
}
\end{document}